\documentclass[12pt]{iopart}
\usepackage{graphicx,times,iopams}

\newcommand{\bra}[1]{\langle #1|}
\newcommand{\ket}[1]{| #1 \rangle}
\newcommand{\op}[1]{\hat{#1}}
\newcommand{\cre}[1]{\hat{#1}^\dagger}
\newcommand{\des}[1]{\hat{#1}}

\newcommand{\osum}{\circlearrowleft\hspace{-1.4em}\sum}

\begin{document}

\title{Synthetic gauge fields and homodyne transmission in Jaynes-Cummings lattices}
\author{A.\ Nunnenkamp$^1$, Jens Koch$^2$, and S.\ M.\ Girvin$^1$}
\address{$^1$ Departments of Physics and Applied Physics, Yale University, New Haven, CT~06520, USA}
\address{$^2$ Department of Physics and Astronomy, Northwestern University, Evanston, IL~60208, USA}
\ead{andreas.nunnenkamp@yale.edu}

\begin{abstract} Many-body physics is traditionally concerned with systems of interacting massive particles. Recent studies of effective interactions between photons, induced in the circuit QED architecture by coupling the microwave field to superconducting qubits, have paved the way for photon-based many-body physics. We derive the magnitude and intrinsic signs of photon hopping amplitudes in such circuit QED arrays. For a finite, ring-shaped Jaynes-Cummings lattice exposed to a synthetic gauge field we show that  degeneracies in the single-excitation spectrum emerge, which can give rise to strong correlations for the interacting system with multiple excitations. We calculate the homodyne transmission for such a device, explain the generalization of vacuum Rabi splittings known for the single-site Jaynes-Cummings model, and identify fingerprints of interactions beyond the linear response regime.
\end{abstract}

\submitto{\NJP}

\maketitle

\section{Introduction}

The notion of a quantum simulator can be traced back to a remark by Richard Feynman \cite{Feynman1982} which was later made more precise by David Deutsch \cite{Deutsch1985} and Seth Lloyd \cite{Lloyd1996}. Since simulating quantum many-body systems on a classical computer is difficult due to the exponential scaling of the Hilbert space dimension with the number of particles, it is intriguing to try and use a well-controllable quantum system with a tunable Hamiltonian to simulate the physics of another quantum system of interest. Over the last decade, this paradigm has motivated many experiments and theoretical proposals, predominantly for setups with ultracold atoms or trapped ions (see for example Ref.~\cite{Buluta2009} for a recent review).

More recently, numerous proposals have emphasized the potential of the circuit QED architecture \cite{Schoelkopf2008} to serve as an interesting solid-state quantum simulator based on polaritons \cite{Hartmann2008}. Indeed, the physics and well-established fabrication techniques for coupled systems of on-chip microwave resonators and superconducting qubits, offer a number of promising properties. These include an immense freedom in engineering lattices with different geometries, dimensionalities, and topologies in a bottom-up approach \cite{Tsomokos2010}. Manipulation and measurement of quantum states at the single-site level are already standard practice in circuit QED experiments. Finally, the controlled coupling to local or global microwave drives also promises new insight into the nonequilibrium physics of strongly correlated polariton systems.

According to theoretical studies \cite{Angelakis2007, Greentree2006, Hartmann2006, Koch2009a, Schmidt2009}, these systems provide access to the superfluid--Mott insulator phase transition in the setting of the Jaynes-Cummings model, with characteristic similarities and differences compared to the more familiar transition in the popular Bose-Hubbard model \cite{Fisher1989}. Theoretical predictions for the physics of finite Jaynes-Cummings arrays in the presence of drive and decay exist \cite{Carusotto2009, Hartmann2010, Liew2010, Lieb2010, Schmidt2010a, Ferretti2010, Bamba2011, Knap2011}, and proposals for introducing synthetic gauge fields \cite{Koch2010, Kamal2011}  may turn Jaynes-Cummings lattices into fruitful ground for studying novel quantum phases of strongly-correlated bosons with broken time-reversal symmetry. Experimental efforts to extend the demonstration of single-site photon blockade \cite{Hoffman2010} to larger arrays are currently underway.

In this article, we study the minimal circuit QED system in which breaking of time-reversal symmetry is observable: a three-site Jaynes-Cummings ring subject to a synthetic gauge field, which induces nontrivial phase factors in the photon hopping terms. Our study is organized as follows. In Section 2, we review the many-body Hamiltonian for a general circuit QED array, and derive magnitude and intrinsic signs of the photon hopping amplitudes. In Section 3, we develop the formalism to calculate linear-response reflection and transmission amplitudes. These quantities play a central role in all circuit QED experiments to date, which routinely probe quantum states by homodyne measurements. It is to be expected that transmission spectra will continue to serve as a fundamental probe in larger Jaynes-Cummings lattices, and the first experiments are currently underway\footnote{Andrew Houck, private communication (2011).}.

Focussing on the finite Jaynes-Cummings ring, we show in Section 4 the presence of degeneracies in the single-particle spectrum for specific values of the gauge-invariant phase sum. For commensurate polariton occupation, even weak interactions lead to strong correlations and generation of highly-entangled Schr\"odinger cat states. In Section 5, we calculate the transmission through such a device. For weak drive strengths, we confirm transmission peaks at frequencies corresponding to the eigenmodes within the 1-excitation manifold. Beyond this linear response regime, polariton interactions become important, and numerical results show features similar to the nonlinear vacuum Rabi splitting of the single-site Jaynes-Cummings model.

\section{Model of circuit QED lattices}

\begin{figure}
\begin{center}
\includegraphics[width=0.9\columnwidth]{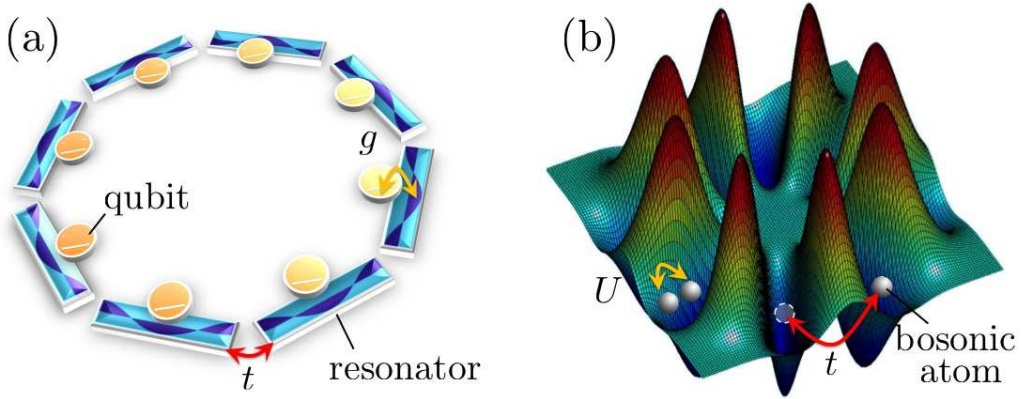}
\caption{(a) Jaynes-Cummings ring lattice: photons hop between resonators with amplitude $t$. Each resonator is coupled locally to a superconducting qubit with strength $g$. (b) Ring lattice for ultracold atoms: bosonic atoms can tunnel between adjacent minima of the trapping potential with amplitude $t$ and feel an on-site interaction of strength $U$.}
\label{fig1}
\end{center}
\end{figure}

In the simplest setting possible, circuit QED arrays comprise a bosonic lattice with two species of lattice sites, cf.\ Fig.\ 1(a). The first species, qubit sites, consists of two-level systems with transition energy $\varepsilon$, realized by suitable superconducting qubits. The second species, resonator sites, describes harmonic oscillators with characteristic energy $\omega$, implemented as a fixed photon mode inside on-chip resonators (e.g.~coplanar waveguides).

Coupling between sites in such a circuit QED lattice allows for photon hopping and conversion between photon and qubit excitations, which in part mimics the hopping of bosonic particles in the Bose-Hubbard model \cite{Fisher1989} and its multi-species generalizations \cite{Altman2003}. The detailed mechanisms for hopping in circuit QED lattices, however, are more diverse: hopping $t$ between two resonator sites corresponds to the transfer of individual photons. As opposed to quantum tunneling, these processes are possible due to classical EM evanescent fields and have a clear correspondence at the classical level. For on-chip microwave resonators, this coupling is commonly realized by capacitors interconnecting two or three resonators. Hopping between resonator and qubit sites $g$ is of the Jaynes-Cummings type, and enables the interconversion of qubit excitations and photons. Direct coupling between qubit sites should be feasible for closely spaced qubits, but, as of now, has not been demonstrated in the circuit QED architecture.

The common situation of small hopping strengths ($t,\,g\ll \omega,\,\epsilon$) affords the rotating-wave approximation, and results in conservation of the total excitation number $\op{N}=\sum_j \cre{a}_j \des{a}_j + \sum_j \op{\sigma}_j^+\op{\sigma}_j^-$, corresponding to the particle number in Bose-Hubbard type models. In total, the Hamiltonian of a circuit QED lattice assumes the form
\begin{equation}\label{Ham0}\fl
\op{H}_\mathrm{lat}=\omega\sum_j \cre{a}_j \des{a}_j + \epsilon \sum_j \op{\sigma}_j^+\op{\sigma}_j^- + t\sum_{\langle j,j' \rangle} (e^{i\theta_{jj'}} \cre{a}_j \des{a}_{j'} +\mathrm{H.c.}) + g\sum_j(\cre{a}_j \op{\sigma}_j^-+\mathrm{H.c.}).
\end{equation}
We note that disorder in the system parameters (beyond the scope of this paper), may need to be included in Eq.~(\ref{Ham0}), once it is characterized experimentally. A lattice topology of particular interest, is that of a finite-size ring, shown in Fig.~\ref{fig1}(a). Its analog for  ultracold atoms, see Fig.~\ref{fig1}(b), has been discussed in the literature \cite{Amico2005, Rey07, Hallwood06a, Nunnenkamp08}, and (if loaded with bosonic atoms) shares some features with the system to be discussed here.

The option to study broken time-reversal symmetry in circuit QED lattices is based on synthetic gauge fields which give rise to complex phases $e^{i \theta_{jj'}}$ in the photon hopping terms. Making photons susceptible to a magnetic field is possible, for example, by substituting inter-resonator coupling capacitances with passive superconducting circuits \cite{Koch2010}. Each such circuit consists of a superconducting loop interrupted by three (or more) Josephson junctions, in which time-reversal symmetry can be broken by an external magnetic flux. Photon hopping takes place via virtual excitations of this coupler circuit, and thus transfers the symmetry breaking to the photonic system. Alternatively, active non-reciprocal devices may be used, as recently pointed out in an interesting proposal by Kamal \etal~\cite{Kamal2011}.

\begin{figure}
\begin{center}
\includegraphics[width=0.95\columnwidth]{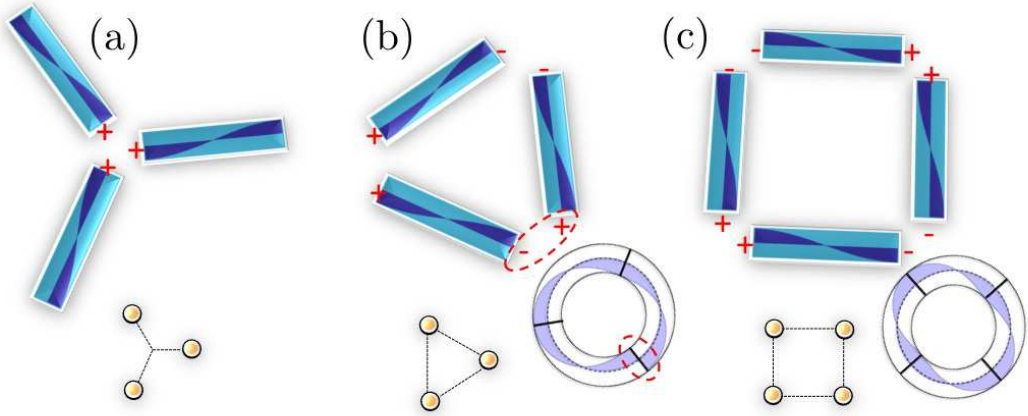}
\caption{\emph{Photon hopping in resonator arrays.} Photon lattices, built from quasi-1d microwave resonators, admit uniform photon hopping due to capacitive coupling between adjacent ends of two or three resonators. When coupling full-wavelength modes (resonator length $L$ matches an integer multiple of the wavelength $\lambda$), photon hopping has a positive sign. For half-wavelength modes, $L=(n+1/2)\lambda$, the mode function has different signs at opposite ends of each resonator, and photon hopping terms  can acquire a gauge-invariant phase of $\pi$. It is important to note that the 1d nature of resonators allows two distinct configurations for coupling 3 resonators to each other, see (a) and (b). Gauge invariant $\pi$ phases only emerge in ring-type coupling (b), and only for rings with an odd number of sites. Rings with even numbers of sites, see (c), do not acquire a $\pi$ phase.}
\label{fig2}
\end{center}
\end{figure}

\subsection{Photon hopping: complex phases and signs} Individually, the phases $e^{i \theta_{jj'}}$  associated with hopping terms are not gauge invariant. Gauge transformations of the type $\des{a}_j\to\des{a}_j e^{i\chi_j}$ lead to a fully equivalent Hamiltonian with altered phase factors  $e^{i\theta_{jj'}+i\chi_{j'}-i\chi_{j}}$. The relevant quantities are gauge-invariant phase sums,
\begin{equation}
\theta_{\Sigma}( \mathcal{C}) =  \osum_{\mathcal{C}[jj']} \theta_{jj'},
\end{equation}
around plaquette paths $\mathcal{C}[jj']$ in the resonator sublattice. If no such plaquettes exist (as is the case, e.g., in an open 1d resonator chain), all phases can be gauged away immediately and the hopping terms may be rewritten with positive, real-valued amplitudes, i.e.\ $t\sum_{\langle j,j' \rangle} (\cre{a}_j \des{a}_{j'} + \mathrm{H.c.})$. For clarity, and to make relevant signs explicit, we will assume $t$ to be real-valued and positive throughout; i.e.~phases and minus signs will not be absorbed into $t$.

As long as time-reversal symmetry is intact, $\theta_\Sigma(\mathcal{C})$ is an integer multiple of $\pi$ on all plaquettes, and hopping amplitudes can be chosen strictly real-valued. This, however, leaves open an important question about the signs of photon hopping terms. Such signs can arise from coupling $\lambda/2$ resonator modes to each other, or, in general, any  half-integer mode with $L=(n+1/2)\lambda$ ($n\in\mathbb{N})$. As illustrated in Fig.\ \ref{fig2},  it is essentially the sign change of the mode function for half-wavelength resonances which introduces a gauge-invariant phase $\theta_\Sigma=\pi$ for each plaquette with an odd number of resonators. 

More specifically, photon hopping between two adjacent resonators $j$ and $j'$ has the amplitude
\begin{equation}
	t\, e^{i\theta_{jj'}}=\frac{1}{2}C_c \omega\, \phi_j(x_\mathrm{end})\, \phi_{j'}(x_\mathrm{end}'),
\end{equation}
where $C_c$ is the coupling capacitance and $\phi_j(x)$ denotes the classical mode function of resonator $j$. For full-wavelength modes, the signs of the mode functions can be chosen to be positive at each coupling point such that hopping amplitudes are positive. Alternatively, for any different sign choice of each mode function $\phi_j$, one can always perform a gauge transformation that renders all hopping amplitudes real and positive.

By contrast, half-wavelength modes are associated with an alternating sign, $\phi_j(x_{\mathrm{end},1}) = -\phi_j(x_{\mathrm{end},2})$, between opposite ends of each resonator, see Fig.\ \ref{fig2}.  The additional signs lead to changes in the energy spectrum or band structure if (and only if) the lattice contains at least one plaquette with an odd number of sites on its boundary. For such odd-number plaquettes, the gauge-invariant phase sum is $\theta_\Sigma (\mathrm{mod}\;{2\pi})=\pi$.

\emph{Derivation of hopping amplitudes.}---We extend the treatment of a single, capacitively coupled resonator \cite{Koch2010} to that appropriate for a full circuit QED lattice, and derive photon hopping amplitudes and their signs in a circuit QED lattice. Based on the usual circuit quantization scheme \cite{Devoret1995}, the Lagrangian for the resonator sublattice is obtained from the LC decomposition of the transmission line resonators and a subsequent continuum limit: 
\begin{equation} \label{lagr}
\mathcal{L}=\frac{1}{2}C_c\sum_j \dot{\Phi}_j^2\bigg|_{\mathrm{ends}}+\frac{c}{2}\sum_j\int dx\bigg[ \dot{\Phi}_j^2-{\textstyle\frac{1}{\ell c}}(\partial_x \Phi_j)^2\bigg] -C_c\sum_{\langle j,j' \rangle} \dot\Phi_j\dot\Phi_{j'}\bigg|_{\mathrm{ends}}.
\end{equation}
Here, $\dot\Phi_j(x,t)$ is the voltage profile in resonator $j$, and $\ell$, $c$ denote inductance and capacitance per unit length. The sum in the last term runs over nearest-neighbor resonators, $C_c$ is the coupling capacitance between each adjacent resonator pair, and the amplitudes are evaluated at the corresponding resonator ends connected to $C_c$. Note that the sign of the coupling term stems from the expansion of the capacitive energy $\frac{1}{2}C_c(\dot\Phi_j - \dot\Phi_{j'})^2|_{\mathrm{ends}}$; diagonal terms of the former expression appear in the Lagrangian (\ref{lagr}), and also affect the boundary conditions for the fields $\Phi_j$ \cite{Koch2010}.

Employing these boundary conditions at the two ports of every resonator, the resonator Lagrangian is diagonalized in terms of internal eigenmodes, $\Phi_j(x,t)=\sum_\mu \zeta_{j\mu}(t) e^{i\omega_\mu t}\phi_{j\mu}(x)$,  and takes the form
\begin{equation}
\mathcal{L}=\frac{1}{2}\sum_{j,\mu}( \dot{\zeta}_{j\mu}^2-\omega_\mu^2\,\zeta^2_{j\mu}) -C_c\sum_{\langle j,j' \rangle}\sum_{\mu,\mu'} \phi_{j\mu}(x)\phi_{j'\mu'}(x') \bigg|_{\mathrm{ends}}\dot\zeta_{j\mu}\dot\zeta_{j'\mu'}.
\end{equation}
In the typical situation of weak coupling between resonators,  $C_c\ll C_\mathrm{res}=c\, L$, we may focus on a single mode, and thus drop mode indices in the following. 

To switch to the Hamiltonian picture, it is convenient to cast the Lagrangian into matrix notation,
$\mathcal{L}=\frac{1}{2}\dot{\vec{\zeta}}\cdot(1-\mathsf{T})\dot{\vec{\zeta}}-\frac{1}{2}\omega^2\vec{\zeta}^2$,
where $\mathsf{T}_{jj'}=\delta_{\langle j,j' \rangle}C_c\phi_j(x)\phi_{j'}(x')|_\mathrm{ends}$ captures the coupling between resonators, and $\delta_{\langle j,j' \rangle}$ denotes the adjacency matrix of the resonator lattice. Legendre transform results in the Hamilton function
\begin{equation}
\mathcal{H}=\frac{1}{2}\vec{\pi}\cdot(1-\mathsf{T})^{-1}\vec{\pi}-\frac{1}{2}\omega^2\vec{\zeta}^2 \quad\approx \quad  \frac{1}{2}\vec{\pi}^2-\frac{1}{2}\omega^2\vec{\zeta}^2 +\frac{1}{2}\vec{\pi}\cdot\mathsf{T}\vec{\pi},
\label{hamarray}
\end{equation}
where the vector $\vec{\pi}$ collects the conjugate momenta $\pi_j = \partial \mathcal{L}/\partial \dot \zeta_j$. The last approximation in Eq.~(\ref{hamarray}) retains the leading order contribution from the resonator coupling. Finally, canonical quantization replaces $\zeta_j\to(\des{a}_j+\cre{a}_j)/\sqrt{2\omega}$, $\pi_j\to(i\cre{a}_j-i\des{a}_j)\sqrt{\omega/2}$, and in conjunction with the rotating-wave approximation, yields the resonator lattice Hamiltonian
\begin{equation}
\op{H}=\omega\sum_j \cre{a}_j\des{a}_j + t \sum_{\langle j,j' \rangle} (\pm)_{jj'}(\cre{a}_j\des{a}_{j'}+\mathrm{H.c.}).
\end{equation}
Expressed in proper frequency units, the photon hopping due to direct capacitive coupling is therefore given by
\begin{equation}
t/2\pi=\frac{1}{2}\nu\, C_c|\phi_j(x)\phi_{j'}(x')|\bigg|_\mathrm{ends}\approx \frac{C_c}{C_\mathrm{res}} \nu,
\end{equation}
where $\nu=\omega/2\pi$ denotes the resonator frequency. As indicated in the discussion above, the signs for photon hopping are  dictated by the product of mode functions involved in the hopping; specifically
\begin{equation}
(\pm)_{jj'}=\mathrm{sign}\; \phi_j(x)\phi_{j'}(x')\bigg|_\mathrm{ends}.
\end{equation}
This result shows that, up to irrelevant sign changes induced by gauge transformations, photon hopping is strictly positive for full-wavelength modes. By contrast, for half-wavelength modes, alternating signs of the mode functions can indeed lead to observable changes. In future experiments, this may offer a convenient method for switching between photon systems with and without frustrated hopping, by a mere doubling of the microwave drive frequency.

\section{Calculation of steady-state transmission rates}

Experimentally, transmission and reflection measurements are by far the simplest probes for circuit QED arrays, and are performed routinely in small circuit QED systems with one or two resonators \cite{Wallraff2004, Schreier2008, Johnson2010}. Conceptually, such measurements are implemented by coupling one resonator (or several) to an external transmission line, through which  microwave photons can be injected,  and subsequently detected in heterodyne or homodyne. For theory, however, the analysis of even this simple setting is challenging for larger numbers of resonators and qubits, since it generally involves both strong correlations, and the complexity of non-equilibrium physics for a driven, damped quantum system.

Assuming that the coupling to the external transmission lines (and any additional baths) is weak, the system may be described by a Lindblad master equation of the form
\begin{equation}\label{master}
\dot{\op{\varrho}} = -i \left[\op{H}', \hat{\varrho} \right] + \kappa\sum_j\mathcal{D}[\des{a}_j] \op\varrho +  \gamma\sum_j\mathcal{D}[{\op{\sigma}}^-_j] \op{\varrho}	+\frac{\gamma_\varphi}{2}\sum_j\mathcal{D}[\op{\sigma}^z_j] \op\varrho. 
\end{equation}
Here, the last three terms involve the damping superoperator $\mathcal{D}[\op{L}]\op\varrho=\op{L}\op{\varrho} \op{L}^\dag-\{\op{L}^\dag \op{L},\op{\varrho}\}/2$ \cite{Alicki2007} and account for photon loss, qubit relaxation and pure dephasing of qubits. For simplicity, we will neglect pure dephasing ($\gamma_\varphi = 0$) and assume decoherence with uniform rates induced by coupling to separate baths. Extensions to shared baths and varying decoherence rates are possible, but can add further complexity to the analysis.

The unitary evolution is captured by the Liouville term, and expressed in terms of the system Hamiltonian in the frame co-rotating with the drive, i.e.\ 
\begin{equation}
\fl \op{H}'=\op{U}\op{H}\op{U}^\dag-i\dot{\op{U}}\op{U}^\dag=\op{H}_\mathrm{lat}-\omega_d\sum_j \cre{a}_j\des{a}_j - \omega_d\sum_j \op{\sigma}^+_j \op{\sigma}^-_j
+\op{H}_\mathrm{d}|_{t=0} = \op{H}_0' +\op{H}_\mathrm{d}|_{t=0}.
\end{equation}
Here, the unitary is $\op{U}=e^{i\omega_d t\,\op{N}}$ and removes the time-dependence of the drive Hamiltonian
\begin{equation}\label{Hdrive}
\op{H}_\mathrm{d}=\sum_j  \left(\Omega_j \, \cre{a}_j e^{-i\omega_d t} + \mathrm{H.c.}\right)
\end{equation}
for a single drive frequency $\omega_d$, and $\op{H} = \op{H}_\mathrm{lat} + \op{H}_\mathrm{d}$ is the Hamiltonian including the drive.

Under steady-state conditions, $\dot{\op{\varrho}}=0$, the master equation (\ref{master}) forms a linear system of equations for the matrix elements of $\op{\varrho}$. Its solution, in principle, enables the prediction of transmission and reflection signals by calculating $\langle \des{a} \rangle = \tr (\des{a}\op{\varrho})$, whose real and imaginary part correspond to the two quadratures of the homodyne signal in transmission. It is important to stress, however, that the brute-force numerical solution of this equation represents a formidable challenge even for arrays with only a handful of resonators and qubits: the number of independent real variables to be determined from $\dot{\op{\varrho}}=0$ is given by $(\Lambda+1)^{2N_\mathrm{r}} 2^{2N_\mathrm{q}}-1$, where $N_r$ and $N_q$ denote the number of resonators and qubits in the lattice, and $\Lambda$ is the cutoff in photon number. As an example: a lattice consisting of 3 resonators and 3 qubits leads to a linear system of equations with $10^5-1$ variables when admitting up to 4 photons in each resonator. A more appropriate cutoff scheme, not based  on a truncation of the photon Hilbert space, will be described in Section 5.

In light of this scaling, it is useful to note that the direct calculation of the full reduced steady-state density matrix can be  avoided, and instead, only those elements of $\op{\varrho}$ be evaluated that indeed contribute to the expectation value $\langle\des{a}\rangle$. To see this, consider the vanishing time derivative of the steady-state expectation value, and re-express it via the master equation:
\begin{equation}\label{aequation}
0=\frac{d}{dt}\langle\des{a}_j\rangle=\tr(\dot{\op{\varrho}}\des{a}_j) = -i\left\langle [\des{a}_j ,  \op{H}_0' ] \right\rangle - i\Omega_j -\frac{\kappa}{2}\langle \des{a}_j \rangle.
\end{equation}
Not surprisingly, this equation for $\langle a_j \rangle$ in general does not close after explicit evaluation of the commutator. For interacting systems it instead produces an infinite hierarchy of equations, involving an ever increasing number of expectation values of additional operators, which is analogous to situation of the equations of motion method for the system's Green's functions. We review the simple and exact solution in linear response first, and discuss numerical results beyond this in Section 5.

Interaction in circuit QED lattices is readily understood as a hard-core repulsion between bosonic excitations on qubit sites, which is so strong that levels beyond the ground and first excited state are pushed to inaccessibly high energies. Formally, the Jaynes-Cummings lattice  thus corresponds to a two-species Bose-Hubbard model with $U_0=0$ on resonator sites and $U_1 \to \infty$ on 2-level sites \cite{Koch2009b}. This interaction, of course, is only effective in the presence of multiple ``particles" (i.e.\ polariton excitations). Hence, it may be ignored in the linear  response regime where weak driving does not lead to multiple excitations. In this limit, all qubit sites can be replaced by a second set of harmonic oscillator sites, $\op{\sigma}^-_j\to \des{b}_j$, and the resulting lattice Hamiltonian, 
\begin{equation}\fl
\op{h}_\mathrm{lat}=\Delta_\omega\sum_j \cre{a}_j\des{a}_j + \Delta_\epsilon\sum_j\cre{b}_j\des{b}_j + t\sum_{\langle j,j' \rangle}(e^{i\theta_{jj'}}\cre{a}_j\des{a}_{j'}+\mathrm{H.c.}) + g\sum_j (\cre{a}_j\des{b}_{j}+\mathrm{H.c.}),
\end{equation}
describes a simple system of coupled oscillators. 

Using $\Delta_\upsilon=\upsilon-\omega_d$ to denote the detuning of the drive from frequency $\upsilon$,
the commutator required for the calculation of $\langle \des{a}_j \rangle$, Eq.\ (\ref{aequation}), produces
\begin{equation}
\left\langle[\des{a}_j ,  \op{h}_\mathrm{lat} ] \right\rangle = \Delta_\omega \langle\des{a}_j\rangle +t\sum_{j'\in \mathrm{nn}(j)}e^{i\theta_{jj'}}\langle \des{a}_{j'} \rangle 
  +g\langle \des{b}_j \rangle,
\end{equation}
where $\mathrm{nn}(j)$ is the set of sites that are nearest neighbor to site $j$. Together with the analogous relations for $d\langle \des{b}_j\rangle/dt$, one obtains the matrix equation
\begin{equation}\label{mateq}
\left( \begin{array}{cc} 
\Delta_\omega-i\frac{\kappa}{2} +\mathsf{T} & g \\
g & \Delta_\epsilon-i\frac{\gamma}{2}
\end{array}\right)
\left( \begin{array}{c} 
 \vec{\alpha}  \\
\vec{\beta} 
\end{array}\right) =
-\left( \begin{array}{c} 
\vec{\Omega}  \\ 0
\end{array}\right).
\end{equation}
Here, we used the following notation: homodyne amplitudes for the two kinds of sites are lumped into the vectors $\vec{\alpha} =(\langle \des{a}_1\rangle,\ldots,\langle \des{a}_n\rangle)$ and $\vec{\beta} =(\langle \des{b}_1\rangle,\ldots,\langle \des{b}_{n'}\rangle)$. Likewise, $\vec{\Omega}=(\Omega_1,\ldots,\Omega_n)$ collects the drive strength for each resonator ($\des{a}_j$ sites). Finally, $\mathsf{T}$ is the  resonator coupling matrix defined by 
$\mathsf{T}_{jj'}=t e^{i\theta_{jj'}}$ if $j'\in\mathrm{nn}(j)$, and $=0$ otherwise. 

From Eq.\ (\ref{mateq}), one obtains the exact solution for the resonator homodyne signals as
\begin{equation}\label{homodyne1}
\vec{\alpha}
=-\bigg(\Delta_\omega  +\mathsf{T} -{\textstyle\frac{g^2}{\Delta_\epsilon-i\gamma/2}}-i{\textstyle\frac{\kappa}{2}}  \bigg)^{-1}\vec{\Omega}.
\end{equation}
For the physical interpretation of this expression, it is appropriate to switch to the eigenmode basis of the resonator subsystem, $\mathsf{U}(\Delta_\omega+\mathsf{T})\mathsf{U}^\dag=\mathrm{diag}(\Delta_{\Upsilon_\nu})$, where $\Delta_{\Upsilon_\nu}$ denotes the drive detuning from eigenmode frequency $\Upsilon_\nu$, and $\mathsf{U}^\dag$ is the matrix of eigenmode amplitudes for the resonators; specifically, $\mathsf{U}_{\nu j}^*=(\vec{\varepsilon}_\nu)_j$ gives the relative amplitude of eigenmode $\nu$ in resonator $j$. Using this basis yields the following expression for the complex homodyne amplitudes
\begin{equation}\label{amp2}
\langle \des{a}_j \rangle = -\sum_{\nu,k}\mathsf{U}_{\nu j}^*
\left( \Delta_{\Upsilon_\nu}-{\textstyle\frac{g^2}{\Delta_\epsilon-i\gamma/2}}-i{\textstyle\frac{\kappa}{2}}\right)^{-1}\mathsf{U}_{\nu k}\, 
\Omega_k.
\end{equation}
The general structure of this expression can be explained as follows. Coherent driving of individual resonators (with amplitudes given by the components of $\vec{\Omega}$) results in an effective drive $\vec{\Omega}_\Upsilon=\mathsf{U}\vec{\Omega}$ of the individual resonator eigenmodes. The complex amplitudes induced in each eigenmode $\nu$  depend on the detuning of the drive frequency from photon modes and qubits as well as the dissipative rates for photon loss and qubit relaxation. Finally, the amplitude measured in resonator $j$ is the weighted sum over all eigenmode amplitudes, with weight factors $\mathsf{U}_{\nu j}^*$ accounting for the overlap between resonator $j$ and eigenmode $\nu$. 

\subsection{Resonator array ($g=0$)}
For $g=0$, transmission only probes the array of coupled resonators. As expected, a drive on a single resonator $k$ results in a transmitted amplitude $T_{jk}(\omega_d)=|\langle \des{a}_j \rangle|$ in resonator $j$, which becomes maximal whenever $\omega_d$ matches one of the eigenfrequencies $\Upsilon_\nu$. The maximum amplitude values (peak heights in the homodyne amplitude) are given by
\begin{equation}\label{Ti0}
T_{jk}(\omega_d=\Upsilon_\nu)=\bigg|\sum_{\nu'} \frac{ \mathsf{U}_{\nu' j}^* \mathsf{U}_{\nu' k}\Omega_k}{ \Delta_{\Upsilon_{\nu'}}-i\kappa/2}\bigg|
\approx \frac{2}{\kappa}\Omega^{(\nu)}_{jk},
\end{equation}
where $\Omega^{(\nu)}_{jk}=|\mathsf{U}_{\nu j}^* \mathsf{U}_{\nu k}\Omega_k|$ denotes the effective drive strength on resonator $j$, mediated by eigenmode $\nu$, and generated by a drive on resonator $k$. The final approximation of Eq.\ (\ref{Ti0}), is valid for non-degenerate and well-separated eigenmodes, i.e.\ $|\Upsilon_{\nu'}-\Upsilon_\nu|\sim t\gg\kappa$. The generalization to degenerate but otherwise well-separated eigenspaces is achieved by including an additional sum over those modes  degenerate with $\nu$.

\subsection{Jaynes-Cummings lattice ($g\not=0$)}
Once the qubit subsystem is coupled to the resonator array, the single-excitation spectrum and corresponding transmission resonances are modified. Focussing again on sufficient peak separation ($g,\,t\gg \kappa,\gamma$), the peak frequencies can be approximated by the poles of Eq.\ (\ref{amp2}) when setting $\kappa=\gamma=0$. This yields 
\begin{equation}\label{eigenfreq-total}
\Theta_{\nu,\pm}={\textstyle\frac{1}{2}(\Upsilon_\nu+ \epsilon)  \pm \frac{1}{2}(\Upsilon_\nu-\epsilon)\sqrt{1+4g^2/(\Upsilon_\nu-\epsilon)^2}}
\end{equation}
as the predicted resonance frequencies for the transmitted amplitude. It is simple to verify that the frequencies $\Theta_{\nu,\pm}$ are identical to the eigenfrequencies of the \emph{full} coupled system of the resonators $\des{a}_j$ and the harmonic modes $\des{b}_j$. The corresponding heights of resonances in the homodyne transmission amplitude are
\begin{equation}\label{Ti}
T_{jk}(\omega_d=\Theta_{\nu,\pm})
\approx \frac{2  (\epsilon -\Theta_{\nu,\pm} )^2}{\kappa  (\epsilon -\Theta_{\nu,\pm} )^2+g^2 \gamma}\Omega^{(\nu)}_{jk}.
\end{equation}
The Jaynes-Cummings physics of resonant vacuum Rabi splitting, and off-resonant dispersive shifts evidently carries over to the single-excitation physics of the Jaynes-Cummings lattice. The one-excitation spectrum, described by the set of frequencies $\{\Theta_{\nu,\pm}\}$, crucially depends on the detunings of the qubit frequency $\epsilon$ from each resonator eigenmode $\Upsilon_\nu$. For each individual eigenmode $\nu$,  detuning can be large, $|\epsilon-\Upsilon_\nu|\gg g$, and render the interaction dispersive, or be near resonant, $|\epsilon-\Upsilon_\nu| \ll g$, and lead to strong photon-qubit hybridization.

\emph{Dispersive regime.}---The dispersive coupling between qubits and resonator mode $\nu$ at large detuning is reflected in small frequency shifts of the qubit frequency $\epsilon$ and the eigenmode $\Upsilon_\nu$. Expansion of Eq.\ (\ref{eigenfreq-total}) to leading order in $g/(\epsilon-\Upsilon_\nu)$ yields the dispersive shifts familiar from the Jaynes-Cummings model,
\begin{equation}
\Theta_{\nu}^\mathrm{photon}\approx  \Upsilon_\nu - \frac{g^2}{\epsilon-\Upsilon_\nu},\qquad \Theta_{\nu}^\mathrm{qubit}\approx \epsilon + \frac{g^2}{\epsilon-\Upsilon_\nu},
\end{equation}
which imply level repulsion between photon and qubit modes.
The corresponding height of the photon resonance is only slightly reduced from its $g=0$ value, whereas the additional qubit resonance is strongly suppressed:
\begin{equation}\fl
T_{jk}(\Theta_\nu^\mathrm{photon})=\frac{2}{\kappa}\left[1-\frac{\gamma}{\kappa}\frac{g^2}{(\epsilon -\Upsilon_{\nu})^2}\right]
\Omega^{(\nu)}_{jk},
\qquad 
T_{jk}(\Theta_\nu^\mathrm{qubit})=\frac{2}{\gamma}\frac{g^2}{(\epsilon -\Upsilon_{\nu} )^2}\Omega^{(\nu)}_{jk}
\end{equation}

\emph{Resonant regime.}---Whenever the qubit frequency is tuned into near degeneracy with one of the photon eigenmodes, their coupling results in hybridized eigenstates and the typical vacuum Rabi splitting. Expansion of equations (\ref{eigenfreq-total}) and (\ref{Ti}) to leading order in $(\epsilon-\Upsilon_\nu)/g$ gives
\begin{equation}\fl
\Theta_{\nu,\pm}=\Upsilon_\nu \pm g \pm (\epsilon -\Upsilon_{\nu} )^2/g,\qquad
T_{jk}(\Theta_{\nu,\pm})=\frac{2}{\gamma+\kappa}\left[1\mp\frac{2\gamma(\epsilon - \Upsilon_\nu)}{g(\gamma+\kappa)}\right]
\end{equation}
for the peak frequencies and heights.

It is interesting to note that this linearization approach, valid for circuit QED arrays at sufficiently small drive powers, may also provide insight beyond the linear response regime for multi-level systems with small anharmonicity. The transmon qubit \cite{Schreier2008, Koch2007a} provides one example for such a system, where under appropriate conditions the anharmonicity may be treated perturbatively \cite{Hartmann2010}.

\section{Strongly-correlated states in a three-site Jaynes-Cummings ring lattice}

\begin{figure}
\begin{center}
\includegraphics[width=0.9\columnwidth]{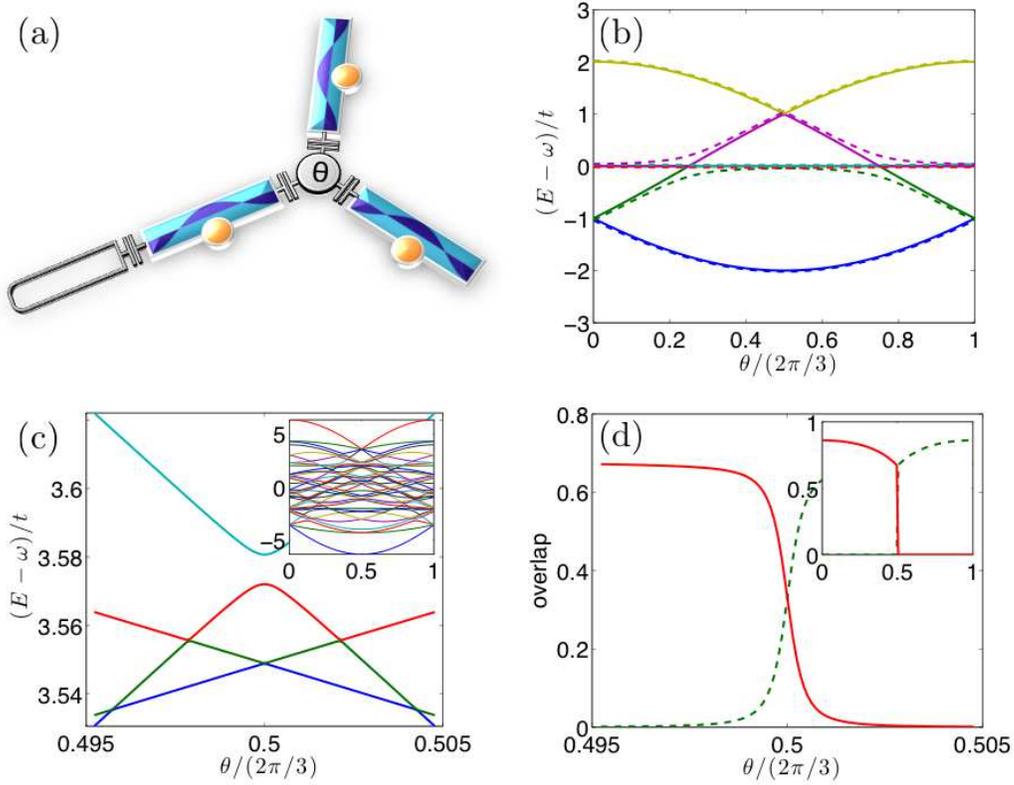}
\caption{\emph{Three-site Jaynes-Cummings ring lattice.} (a) Three resonators are coupled via a Josephson ring circuit, and each resonator interacts with its local superconducting qubit. In Section 5 we will consider the case where one of the resonators is driven by a microwave tone. (b) Spectrum in the 1-excitation manifold for $g=0$ (solid) and $g/t = 0.2$ (dashed). For $\theta = \pi/3$ the highest-energy state is doubly degenerate. (c) Spectrum in the 3-excitation manifold. The degeneracy in the 1-excitation manifold at $\theta = \pi/3$ induces a 4-fold degeneracy in the 3-excitation manifold which is lifted for $g/t = 0.5$. (d) Overlap of the highest-energy 3-excitation state with $\ket{3,0,0}\ket{\downarrow, \downarrow, \downarrow}$ (red solid) and $\ket{0,3,0}\ket{\downarrow, \downarrow, \downarrow}$ (green dashed) as a function of phase twist $\theta$ for $g/t = 0.5$. Around $\theta = \pi/3$ the state has large overlap with the superposition states $\ket{\psi_+}$ in Eq.~(\ref{cat}). The insets in (c) and (d) show zoom-outs of the same graphs. In all cases, the resonators and qubits are in resonance with each other $\varepsilon = \omega$.}
\label{fig3}
\end{center}
\end{figure}

The discussion in Sections 2 and 3 has been general. In the following, we will focus on the concrete example of a three-site Jaynes-Cummings ring lattice, see Fig.~\ref{fig3}(a). The system consists of three resonators coupled via a Josephson ring circuit \cite{Koch2010}, and each resonator interacts with its local superconducting qubit. It is the simplest system for which a gauge field can lead to measurable effects as the nonzero phase sum around the loop cannot be gauged away. In this section we show  how strongly-correlated states can arise even in this simple setup.

Following Section 2, the Hamiltonian of the three-site Jaynes-Cummings ring lattice is given by
\begin{equation}\label{Ham}
\fl \op{H}_0 = \sum_{j=1}^3 \left[ \omega \cre{a}_j \des{a}_j + \epsilon \op{\sigma}_j^+\op{\sigma}_j^- + g (\cre{a}_j \op{\sigma}_j^-+\mathrm{H.c.}) \right] + t \left[ e^{i\theta} (\cre{a}_2 \des{a}_1 + \cre{a}_3 \des{a}_2 + \cre{a}_1 \des{a}_3) + \mathrm{H.c.}) \right]
\end{equation}
where $\theta = \theta_\Sigma/3$ denotes the phase twist and $\theta_\Sigma$ is the gauge-invariant phase sum. We note that a similar Hamiltonian arises for bosons in rotating ring lattices \cite{Rey07, Hallwood06a, Nunnenkamp08}.

Since $\op{H}_0$ (\ref{Ham}) conserves the total excitation number $\op{N} = \sum_{j=1}^3 \left( \cre{a}_j \des{a}_j + \op{\sigma}_j^+ \op{\sigma}_j^- \right)$, its energy spectrum and eigenfunctions can be studied for each $N$ separately. In the subspace with one excitation ($N=1$) we can regard the qubits as a second class of harmonic oscillators: $\des{\sigma}_j \rightarrow \des{b}_j$. Since the system has translational symmetry it is advantageous to introduce quasi-momentum operators $\des{a}_j = \frac{1}{\sqrt{3}} \sum_{k=0}^2 e^{-2\pi i k j/3} \des{a}_k$ and $\des{b}_j = \frac{1}{\sqrt{3}} \sum_{k=0}^2 e^{-2\pi i k j/3} \des{b}_k$. These reduce the Hamiltonian to the form
\begin{equation}
\op{H}_0^{(1)}
= \sum_{k=0}^{2}
\left( \begin{array}{cc} \cre{a}_k & \cre{b}_{k} \end{array}\right)
\left( \begin{array}{cc}
\varepsilon_k & g \\
g & \epsilon
\end{array} \right)
\left( \begin{array}{c} \des{a}_k \\ \des{b}_{k} \end{array}\right)
\end{equation}
with the eigenmode frequencies $\Upsilon_k = \omega+2t \cos\left(\frac{2\pi k}{3}+\theta\right)$ of the coupled resonator sublattice, and eigenvalues
\begin{equation}
\Theta_k^\pm = \frac{\Upsilon_k + \epsilon}{2} \pm \frac{\sqrt{ (\Upsilon_k - \epsilon)^2 + 4 g^2}}{2}.
\end{equation}

Fig.~\ref{fig3}(b) shows the 1-excitation energy spectrum as a function of the phase twist $\theta$. For $g=0$, the spectrum consists of the resonator modes $\Upsilon_k$ and the three uncoupled qubits at $\epsilon =\omega$. For $g \not= 0$ several level crossings between qubit and photon excitations turn into avoided crossings, but at the critical phase twist $\theta = 0$, the ground state (and for $\theta = \pi/3$, the highest-energy state) in the 1-excitation manifold remains doubly degenerate, i.e.~$\Theta_1^- = \Theta_2^-$ for $\theta = 0$ and $\Theta_0^+ = \Theta_1^+$ for $\theta = \pi/3$. Accordingly,  even without synthetic gauge field, $\theta = 0$, the plaquette is frustrated. The degeneracy in the 1-excitation spectrum  in turn leads to strong correlations among the photons even for weak interaction strength, as we will show next.

For $g=0$ each twofold degeneracy in the 1-excitation spectrum induces an ($N+1$)-fold degeneracy in the $N$-excitation spectrum. Focussing on the critical phase twist $\theta = \pi/3$, this subspace is spanned by $\ket{n_{k=0}=n,n_{k=1}=N-n,n_{k=-1}=0} \ket{\downarrow,\downarrow,\downarrow}$ with $0 \le n \le N$. Here, $n_k$ denotes the number of excitations with quasi-momentum number $k$.

If (and only if) the number of lattice sites $L$ and the number of excitations $N$ are commensurate, these degeneracies are lifted by the Jaynes-Cummings term for $g \not= 0$. To understand the induced correlations, we rewrite the Hamiltonian (\ref{Ham}) as
\begin{eqnarray}
\hat{H}_0 = \sum_{k=0}^2 \Upsilon_k \cre{a}_k \des{a}_k
+ \epsilon \sum_{j=1}^3 \op{\sigma}_j^+\op{\sigma}_j^- + g \sum_{k,j} \left( \frac{e^{2\pi i j k/3}}{\sqrt{3}}\hat{\sigma}^-_j \cre{a}_k + \mathrm{H.c.} \right).
\end{eqnarray}
The last term lifts the degeneracy between $\ket{N,0,0}\ket{\downarrow, \downarrow, \downarrow}$ and $\ket{0,N,0}\ket{\downarrow, \downarrow, \downarrow}$ in a $2N$-order process, and the new highest-energy eigenstates are the cat-like superposition states
\begin{equation}\label{cat}
\ket{\psi_\pm} = \frac{\ket{N,0,0} \pm \ket{0,N,0}}{\sqrt{2}}\ket{\downarrow, \downarrow, \downarrow}.
\end{equation}

Figures \ref{fig3}(c) and (d) show the spectrum in the 3-excitation manifold ($g/t = 0.5$, $\epsilon = \omega$) as a function of phase twist, and the corresponding overlaps of the highest-energy eigenstate with the states $\ket{3,0,0}\ket{\downarrow, \downarrow, \downarrow}$ and $\ket{0,3,0}\ket{\downarrow, \downarrow, \downarrow}$. At phase twist $0 \le \theta \ll \pi/3$ the highest-energy state has a large overlap with the state $\ket{3,0,0} \ket{\downarrow, \downarrow, \downarrow}$. Close to the critical phase twist $\theta = \pi/3$, four energy levels approach each other. This is the ($N+1$)-fold degenerate subspace in the $g=0$ system. For $g \not= 0$ the degeneracy is lifted and the two highest-energy states have a large overlap with the cat-like superpositions $\ket{\psi_\pm}$. For phase twists $\pi/3 \ll \theta \le 2\pi/3$, the highest-energy state has a large overlap with the state $\ket{0,3,0} \ket{\downarrow, \downarrow, \downarrow}$.

We see that synthetic gauge fields can lead to degeneracies in the 1-excitation spectrum which may induce strong correlations in the interacting many-excitation system.

\section{Transmission of a three-site Jaynes-Cummings ring lattice}

\begin{figure}
\begin{center}
\includegraphics[width=1.0\columnwidth]{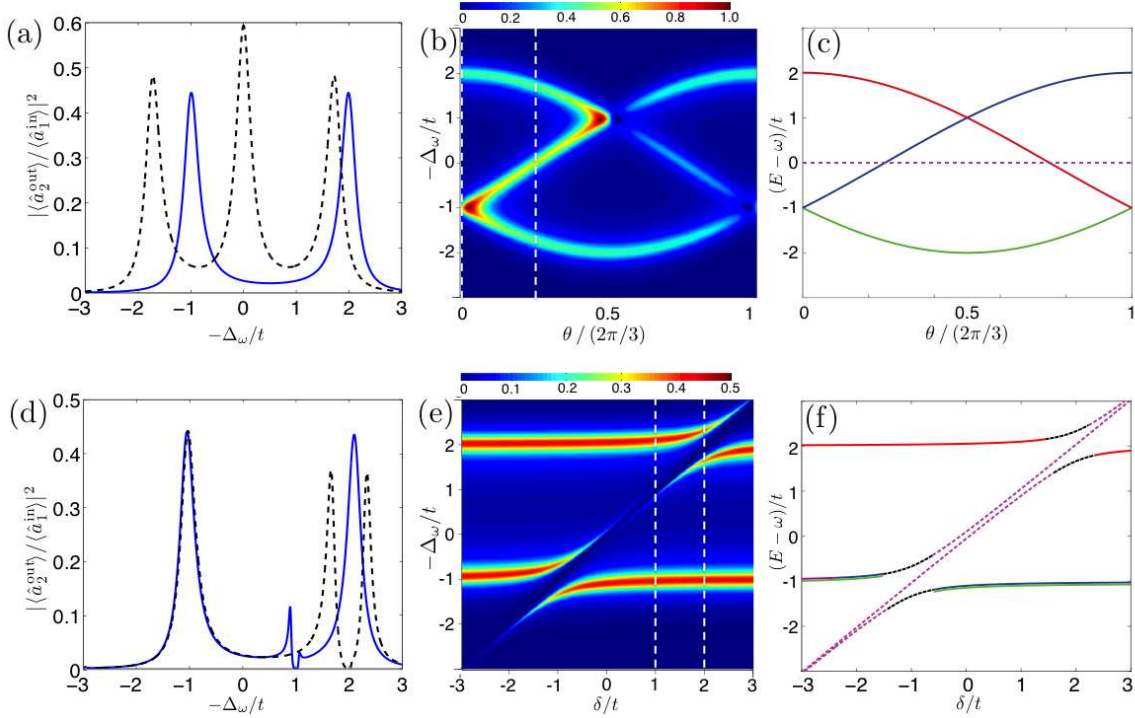}
\caption{\emph{Transmission in linear response.} (a), (b), (d) and (e) show the transmission at port 2, $|\langle \des{a}_2^\mathrm{out} \rangle|^2$, in response to driving at port 1, $|\langle \des{a}_1^\mathrm{in} \rangle|^2$: in (a) for $g = 0$ as a function of drive detuning for $\Delta_\omega$ for $\theta = 0$ (solid) and $\theta = \pi/6$ (dashed) and in (d) for $g/\kappa = 1$ as well as $\delta/t = 1$ (solid) and $\delta/t = 2$ (dashed). Transmission for $g = 0$ as a function of phase twist $\theta$ and drive detuning $\Delta_\omega$ is shown in (b) and for $g/\kappa = 1$ as a function of detuning $\delta$ and drive detuning $\Delta_\omega$ in (e). For comparison, we plot the spectrum of the 1-excitation manifold as a function of phase twist $\theta$ for $\delta = g = 0$ in (c) and as a function of detuning $\delta$ for $g/t = 1/3$ and $\theta = 0$ in (f). The other parameters are $t/\kappa = 3$ and $\gamma/\kappa = 0.1$. In (c) and (f) resonator-like and qubit-like eigenstates are indicated by solid and dashed lines, respectively.}
\label{fig4}
\end{center}
\end{figure}

To establish contact with experimental measurements of homodyne transmission, we apply the general discussion of Section 3 to the finite Jaynes-Cummings ring. As indicated in Fig.~\ref{fig3}(a), we consider driving a single resonator with a microwave tone of strength $\Omega$. All resonators and qubits are subject to dissipation leading to line widths $\kappa$ and $\gamma$, respectively.

As mentioned above, the eigenmodes of the resonator array are characterized by the quasi-momentum number $k$, and the amplitude of eigenmode $k$ in resonator $j$ is $\mathsf{U}_{k j}^*=(\vec{\varepsilon}_k)_j = e^{-2\pi i k j/L}/\sqrt{L}$, independent of phase twist $\theta$.
Using the input-output relations $\hat{a}_j^\mathrm{out} = \hat{a}_j^\mathrm{in} + \sqrt{\kappa_j} \hat{a}_j$ and $\sqrt{\kappa_j} \langle \hat{a}_j^\mathrm{in} \rangle = i \Omega_j$ \cite{GirvinRMP}, we obtain the reflection and transmission rates from the expectation values of the photon annihilation operators,  $\langle \hat{a}_j \rangle$. In reflection, the drive interferes with the reflected wave from the cavity, i.e.
\begin{equation}
\left|\frac{\langle \hat{a}_1^\mathrm{out} \rangle}{\langle \hat{a}_1^\mathrm{in} \rangle}\right|^2 = \left|1 - \frac{\langle \hat{a}_1 \rangle}{i\Omega/\kappa}\right|^2
\hspace{1cm} \textrm{and} \hspace{1cm}
\left|\frac{\langle \hat{a}_j^\mathrm{out} \rangle}{\langle \hat{a}_1^\mathrm{in} \rangle}\right|^2 = \left|\frac{\langle \hat{a}_j \rangle}{\Omega/\kappa}\right|^2 = \frac{\left|\langle \hat{a}_j \rangle\right|^2}{\Omega^2/\kappa^2}.
\end{equation}
In the absence of coupling between resonator and qubits ($g = 0$) or if qubit dissipation is negligible ($\gamma = 0$), we have $\sum_j  \langle \hat{n}_j^\mathrm{out} \rangle = \langle \hat{n}_1^\mathrm{in} \rangle$ expressing the conservation of the photon flux $\hat{n}_j^\mathrm{in/out} = (\cre{a}_j)^\mathrm{in/out} (\des{a}_j)^\mathrm{in/out}$.

We first consider the case $g = 0$, where qubits and resonators are decoupled. Fig.~\ref{fig4}(a) shows the transmission  $|\langle \des{a}_2 \rangle|^2$ as a function of detuning between resonator and drive frequency $\Delta_\omega$. As expected, peaks in transmission correspond to the eigenfrequencies of the resonator array. For $\theta = 0$ the quasi-momentum states $k =\pm 1$ are degenerate leading to only two peaks at $-\Delta_\omega/t = -1$ and $-\Delta_\omega/t = +2$. For $\theta \not= 0$ the spectrum is in general non-degenerate and the three peaks mark the three quasi-momentum modes of the ring lattice. In Fig.~\ref{fig4}(b), the dependence of transmission on drive detuning $\Delta_\omega$ and phase twist $\theta$ shows that the transmission  follows the resonator dispersion $\varepsilon_k(\theta)$, cf.\ Fig.~\ref{fig4}(c). An interesting detail: when the spectrum is almost degenerate  transmission is very sensitive to changes in the phase twist, see, e.g., the situation of $-\Delta_\omega/t = 1$ close to $\theta = \pi/3$. For these parameters the circuit, in fact, implements an on-chip circulator (different from the parameter choice of Ref.\ \cite{Koch2010}).

For $g \not= 0$ and small drive, linear response theory remains valid and qubit sites can be replaced by harmonic oscillators as discussed in Section 3. Results from this treatment are shown in Fig.~\ref{fig4}(d) where the transmission $|\langle \des{a}_2 \rangle|^2$ is plotted as a function of detuning $\Delta_\omega$ ($g/t = 1$ and fixed $\theta = 0$). For $\delta = \epsilon - \omega = t$, the qubits are not in resonance with any of the eigenmodes of the resonator array. Relative to the non-interacting case $g=0$, the coupling to the qubits slightly shifts the frequency of the resonator peaks and three additional peaks (two of them degenerate at $\theta = 0$) appear in the spectrum close to $-\Delta_\omega = \delta$. By contrast, for $\delta/t = 2$ the qubits are in resonance with the $k = 0$ resonator mode and, analogous to the vacuum Rabi splitting in the single-site Jaynes-Cummings model, the resonance peak splits. In Fig.~\ref{fig4}(e) we show the transmission $|\langle \des{a}_2 \rangle|^2$ as a function of the detuning $\delta$ between qubit and resonator and drive detuning $\Delta_\omega$. As the qubit frequency $\epsilon$ is varied, qubits are tuned in and out of resonance with each of the resonator modes. For comparison, Fig.~\ref{fig4}(f) depicts the spectrum of the system in the 1-excitation manifold as a function of qubit detuning $\delta$.

For stronger drive $\Omega$, population beyond the 1-excitation subspace sets in, and linear response theory breaks down. To find the transmission in that case, we compute the relevant expectation values via the steady-state solution of the full quantum master equation (\ref{master}). As mentioned above, even  small resonator arrays and moderate photon number cutoffs lead to matrix sizes for the superoperator which quickly become impractical. This problem can be mitigated slightly by an adaptive truncation scheme for the density matrix elements. First, note that the Hamiltonian $\hat{H}_0$ conserves the number of excitations $N$, while the drive Hamiltonian $\hat{H}_d$ induces transitions between subspaces of different excitation numbers. Each element of the standard Fock basis $\ket{\mu}$ is associated with an excitation number $N$. Now, assigning to each element of the density matrix $\varrho_{\mu \nu} = \bra{\mu} \hat{\varrho} \ket{\nu}$ a generalized excitation number $\tilde{N}$ defined as the sum of the excitation numbers of the states $\ket{\mu}$ and $\ket{\nu}$, a truncation based on $\tilde{N}$ is possible where the cutoff is adjusted with increasing drive strength $\Omega$.  This cutoff scheme is more appropriate and fine-grained compared to a maximum photon number cutoff in Fock space.

\begin{figure}
\begin{center}
\includegraphics[width=\columnwidth]{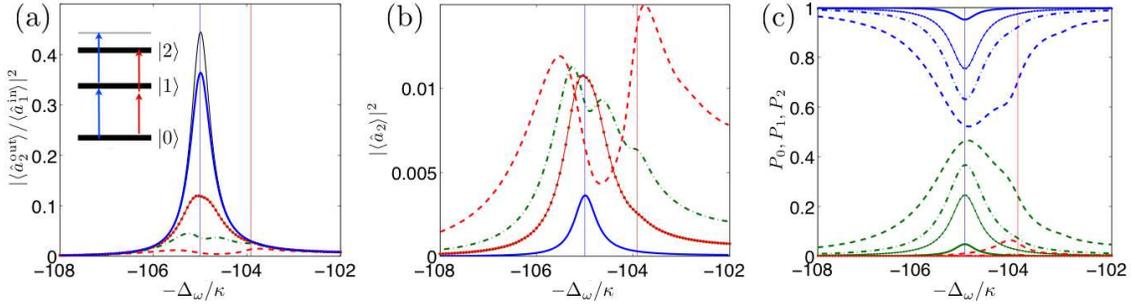}
\caption{\emph{Transmisson beyond linear response.} (a) Transmission $|\langle \des{a}_2^\mathrm{out} \rangle /\langle \des{a}_1^\mathrm{in} \rangle|^2$ and (b) $|\langle \des{a}_2 \rangle|^2$ as a function of detuning $\Delta_\omega$. Linear response approximation (thin black solid in (a)) and numerical solution to quantum master equation (\ref{master}) with cutoff $\tilde{N} = 4$ for $\Omega/\kappa = 0.1$ (blue solid), $\Omega/\kappa = 0.3$ (brown solid with dots), $\Omega/\kappa = 0.5$ (green dash-dotted), and $\Omega/\kappa = 1$ (red dashed). Other parameters are $t/\kappa = 5$, $g/\kappa = 100$, $\delta/\kappa = 5$, and $\gamma = \theta = 0$. Blue and red vertical lines indicate the one-photon and one half of the two-photon transition energies, respectively. (c) For various driving strengths, probability $P_0$ (blue), $P_1$ (green), and $P_2$ (red) to have zero, one, and two excitations in the system. Inset in (a) illustrates the fact that the 2-photon transition energy $\ket{0} \rightarrow \ket{2}$ is detuned from twice the 1-photon transition $\ket{0} \rightarrow \ket{1}$.}
\label{fig5}
\end{center}
\end{figure}

In Fig.~\ref{fig5}(a) and (b) we present the transmission $|\langle \des{a}_2 \rangle|^2$ as a function of drive detuning $\Delta_\omega$ for different drive strengths $\Omega/\kappa$ (cutoff: $\tilde{N} = 4$). For weak drive, the response is close to the linear-response result. As the drive increases, the transmission is suppressed compared to the linear response value. As better seen in Fig.~\ref{fig5}(b), the peak additionally becomes asymmetric, shifts to smaller detunings and finally at strong drive splits into two. In Fig.~\ref{fig5} we have marked the 1-photon transition energies and half of the 2-photon transition energies, revealing that the second peak appears at half the 2-photon transition frequency. This is further corroborated in Fig.~\ref{fig5}(c) where the probabilities for zero, one and two excitations are plotted. At the frequency of the 1-photon transition increasingly strong driving leads to a larger population of the 1-excitation sector $P_1$ but the probability for two excitations $P_2$ remains small. At half the 2-photon transition frequency the population of the 1-excitation space $P_1$ remains small, but at stronger drive strength the 2-excitation manifold is occupied.

This behavior resembles the nonlinear vacuum Rabi splitting as observed in a single-site Jaynes-Cummings system \cite{Bishop2008}. Since the spectrum is nonlinear, photon blockade \cite{Hoffman2010} prevents population transfer from the 1-excitation to the 2-excitation subspace due to the energy mismatch of $E_1-E_0$ vs.\ $E_2-E_1$, see inset of Fig.\ref{fig5}(a). Specifically, driving the system on the 1-photon transition $\ket{0} \rightarrow \ket{1}$ transfers population from the ground state $\ket{0}$ to the first excited state $\ket{1}$, but a second excitation with the same energy cannot enter the system. On the other hand, driving the system at half the 2-photon transition frequency $\ket{0} \rightarrow \ket{2}$, not very effective at low power, allows for two-photon transitions at sufficiently large power. The features of the transmission signal can thus be explained naturally as a consequence of the nonlinear spectrum of the system.

\section{Conclusion}

In summary, we have explored the physics of a Jaynes-Cummings ring, the minimal circuit QED system manifesting effects of broken time-reversal symmetry. Analogous to bosons in rotating ring lattices, degeneracies at specific values of the gauge field and interactions lead to strongly-correlated states. As future experiments are likely to employ transmission as one probe for such systems, we have calculated the response to drive and decay, generalizing phenomena known from the single-site Jaynes-Cummings model, e.g.~the vacuum Rabi and nonlinear vacuum Rabi splittings.
It is clear that a theoretical description of transmission experiments with larger lattices and/or higher drive power will require novel simulation techniques beyond the explicit calculation of the steady-state density matrix of the master equation. Analytical and numerical methods from the field of quantum transport may turn out to be adaptable to this novel setting.
Other valuable experimental probes beyond transmission, such as two-tone spectroscopy and correlation functions,  reveal  additional information about the system, and will be explored in future work.

\ack We acknowledge support from NSF under Grant No.~DMR-1004406 and DMR-0653377 (AN and SMG). Part of the calculations were performed with the Quantum Optics Toolbox \cite{Tan1999}.


\section*{References}

\providecommand{\newblock}{}

\end{document}